\documentclass[twocolumn,showpacs,prb,preprintnumbers,amsmath,amssymb]{revtex4}
\makeatletter
\def\@dotsep{4,5}
\makeatother

\usepackage{graphicx}
\usepackage{dcolumn}
\usepackage{bm}
\usepackage{subfigure} 

\begin{document}

\preprint{}

\title{Scaling properties of critical bubble of homogeneous nucleation in stretched fluid of square-gradient density-functional model with triple-parabolic free energy}

\author{Masao Iwamatsu}
\email{iwamatsu@ph.ns.musashi-tech.ac.jp}
\affiliation{
Department of Physics, General Education Center,
Musashi Institute of Technology,
Setagaya-ku, Tokyo 158-8557, Japan
}%


\date{\today}

\begin{abstract}
The square-gradient density-functional model with triple-parabolic free energy is used to study homogeneous bubble nucleation in a stretched liquid to check the scaling rule for the work of formation of the critical bubble as a function of scaled undersaturation $\Delta\mu/\Delta\mu_{\rm spin}$, the difference in chemical potential $\Delta\mu$ between the bulk undersaturated and saturated liquid divided by $\Delta\mu_{\rm spin}$ between the liquid spinodal and saturated liquid.  In contrast to our study, a similar density-functional study for a Lennard-Jones liquid by Shen and Debenedetti [J. Chem. Phys. {\bf 114}, 4149 (2001)] found that not only the work of formation but other various quantities related to the critical bubble show the scaling rule, however, we found virtually no scaling relationships in our model near the coexistence.  Although some quantities show almost perfect scaling relations near the spinodal, the work of formation divided by the value deduced from the classical nucleation theory shows no scaling in this model even though it correctly vanishes at the spinodal.  Furthermore, the critical bubble does not show any anomaly near the spinodal as predicted many years ago.  In particular, our model does not show diverging interfacial width at the spinodal, which is due to the fact that compressibility remains finite until the spinodal is reached in our parabolic models.     
\end{abstract}

\maketitle

\section{Introduction}
\label{sec:level1}
Homogeneous bubble nucleation (cavitation) has attracted much attention for many years from a fundamental point of view as well as for technological interests.  It occurs when a fluid is undersaturated and is held at a pressure lower than its coexisting vapor pressure at a given temperature.  Then the fluid is said to be stretched.  Homogeneous bubble nucleation occurs in the stretched liquid.  Homogeneous nucleation including homogeneous bubble nucleation in the liquid~\cite{Debenedetti} as well as liquid droplet condensation\cite{Oxtoby} or the crystallization~\cite{Kelton} from melts or glass in general is the basic mechanism to initiate the first-order phase transition, and has been studied for many years mainly using a very simple analytical theory called classical nucleation theory (CNT)~\cite{Debenedetti,Kelton,Oxtoby}.  

Recently, due to the development of the density functional theory~\cite{Debenedetti,Oxtoby} the deviations from the CNT has been predicted using various model systems.~\cite{Oxtoby2,Zeng,Iwamatsu1,Bagdassarian}.  These predictions have successfully explained the deviation of the experimental data from the CNT predictions~\cite{Debenedetti,Oxtoby}.  Therefore, a reliable yet handy empirical formula which can predict nucleation rates is highly desired.  To this end, various empirical formulae for bubble and droplet nucleation in a fluid~\cite{McGraw,Talanquer,Kashchiev,Kusaka,Kalikmanov} as well as crystal nucleation in melts~\cite{Iwamatsu2,Iwamatsu3} have been proposed.

The central quantities of the nucleation phenomena is the nucleation rate $J$, which is the number of critical nuclei formed per unit time per unit volume.  Usually it is written in Arrhenius form
\begin{equation}
J = A \exp\left(-\frac{W^{*}}{k_{\rm B}T}\right)
\label{eq:0}
\end{equation}
where $A$ is a kinetic pre-exponential factor which is believed to be weakly dependent on temperature $T$, $k_{\rm B}$ is the Boltzmann's constant, and $W^{*}$ is the reversible work of formation of the critical bubble. Therefore, the temperature dependence of the nucleation rate is controlled by the work of formation $W^{*}$, which most of the theoretical works study.      

The density functional theory is a powerful tool to study the density profile and the work of formation $W^{*}$ of critical nucleus.  For example, Oxtoby and Evans~\cite{Oxtoby2} has studied the work of formation $W^{*}$ and the density profile of critical droplets and bubbles of the Yukawa fluid using the so-called non-local density functional theory.  Zeng and Oxtoby~\cite{Zeng} studied the critical nucleus of the Lennard-Jones fluid using the non-local density functional theory.  The present author~\cite{Iwamatsu1} used the local square-gradient density functional theory~\cite{Yang,Evans,Falls} combined with the double-parabolic free energy, and demonstrated that the results of Oxtoby and coworkers~\cite{Oxtoby2,Zeng} can be reproducible by the square-gradient density functional theory rather than fully non-local theory.  A similar model with triple-parabolic free energy is used later by Barrett~\cite{Barrett}.  Very recently, Li and Wilemsky~\cite{Li} has compared the results obtained from the accurate non-local density functional theory with that from the approximate square-gradient theory and found that the two results agree qualitatively well.

Among various nucleation processes, bubble nucleation has attracted relatively little attention~\cite{Oxtoby2,Zeng,Iwamatsu1,Muller}.  Recently, Shen and Debenedetti~\cite{Shen} studied homogeneous bubble nucleation in a stretched Lennard-Jones fluid using the non-local density functional theory of Zeng and Oxtoby~\cite{Zeng}.  They found that if scaled by appropriate quantities, various quantities of the critical bubble including the work of formation $W^{*}$ degenerate into universal curves when they are plotted against the scaled undersaturation $\Delta\mu/\Delta\mu_{\rm spin}$.  Their results justify various empirical scaling approach~\cite{McGraw,Talanquer,Kashchiev,Kusaka,Kalikmanov} to the nucleation rate.  Similar density functional calculations have been conducted by Kusaka not only for the Lennard-Jones fluid but for the square-well fluid.  They found that the scaling is modelatedly successfull for those two fluid systems. Recently, Punnathanam and Corti~\cite{Punnathanam1,Punnathanam2} studied the cavity formation rather than the bubble formation and found a similar scaling rule.

All those studies mentioned above are based on the numerical results for several specific models of simple fluids.  However, since the nucleation is so general and ubiquitous phenomena not restricted to simple fluid that it is highly desirable to study the scaling rule in more general models which can be applied not only to the simple liquid but to complex fluids, or even to amorphous or liquid metals.

In this paper, we will use a simple square-gradient density functional theory with a triple-parabolic free energy proposed by Gr\'an\'asy and Oxtoby~\cite{Granasy1} to study the various properties of the critical bubble of homogeneous bubble nucleation.  We choose this model as it captures the most basic properties of nucleation, yet many physical quantities can be handled analytically.  In this study we particularly focus on the scaling rule of various quantities~\cite{Shen}. In Section II we present a short review of the square-gradient density-functional model with triple-parabolic free energy~\cite{Granasy1} to summarize the necessary formula.  We also correct a few typographical errors in the original article~\cite{Granasy1}.  In Section III, we will present the numerical results and discuss the implication of the results in light of the scaling rule.  Finally Section IV is devoted to the concluding summation.

\section{Square-gradient density-functional Model}
\subsection{Triple-Parabolic Free Energy}
In the square-gradient density-functional model of the fluid~\cite{Yang,Evans,Falls}, the free energy of the inhomogeneous fluid, such as the critical bubble in the stretched liquid is written as
\begin{equation}
W=\int \left(\Delta\omega(\phi)+c\left(\nabla \phi\right)^{2}\right) d^{3}{\bf r}.
\label{eq:1}
\end{equation}
This form of the free energy is also known as the Cahn-Hilliard model~\cite{Cahn1,Cahn2} or the phase-field model~\cite{Castro,Granasy2}.   In the triple-parabola model of Gr\'an\'asy and Oxtoby~\cite{Granasy1}, the local part of the free energy $\Delta\omega$ is given by
\begin{equation}
\Delta \omega(\phi) = \left\{
\begin{array}{ll}
\frac{\lambda_{0}}{2}\left(\phi-\phi_{0}\right)^{2}+\Delta\mu\;\; & \phi<\phi_{A} \\
\frac{\lambda_{1}}{2}\left(\phi-\phi_{1}\right)^{2}-\Delta\mu\frac{\phi_{1}-\phi_{0}}{\phi_{2}-\phi_{0}}+\Delta\mu+\omega_{0} & \\
& \phi_{A}\leq\phi \leq\phi_{B} \\
\frac{\lambda_{2}}{2}\left(\phi-\phi_{2}\right)^{2}\;\; & \phi_{B}<\phi
\end{array}
\right.
\label{eq:2}
\end{equation}
which consists of three parabola centered at the vapor density $\phi_{0}$, and at the free energy barrier $\phi_{1}$ (spinodal), and at the liquid density $\phi_{2}$, which we call "vapor", "spinodal" and "liquid" part of the free energy.  We have chosen this triple-parabolic model~\cite{Granasy1} rather than a simpler double-parabolic model~\cite{Iwamatsu1} as the former seems more realistic near the spinodal point.   

Although Gr\'an\'asy and Oxtoby~\cite{Granasy1} set $\phi_{0}=0$ and $\phi_{2}=1$ to simplify the various formulae, we will leave two quantities finite as we want to consider the vapor phase with finite density $\phi_{0}$.  The curvature of parabola $\lambda_{0}$ and $\lambda_{2}$ are related to the compressibility of vapor and liquid phases, and $\Delta\mu$ is the  free energy difference between the liquid and the vapor.  Although $\Delta\mu$ represents in fact the pressure difference $\Delta P$ as Eq.~(\ref{eq:1}) is the grand potential of open system~\cite{Oxtoby}, we call $\Delta\mu$ chemical potential to make the comparison to the previous work~\cite{Shen,McGraw,Talanquer,Kashchiev,Kusaka,Kalikmanov,Punnathanam1,Punnathanam2} easier since $\Delta P$ is proportional to $\Delta\mu$ through $\Delta P=\rho_{\rm l}\Delta\mu$ with $\rho_{\rm l}$ being the number density of the liquid~\cite{Oxtoby,Shen}.  We use the terminology "over-saturation" when $\Delta\mu$ is positive and "undersaturation" when $\Delta\mu$ is negative.  The stretched liquid in this study corresponds to the under-saturated liquid.  Therefore we will be mainly concerned with the situation when $\Delta\mu<0$.

From the continuity of the free energy $\Delta \omega(\phi)$ at the boundary $\phi_{A}$ and $\phi_{B}$, they are given by
\begin{eqnarray}
\phi_{A}&=&\frac{\lambda_{0}\phi_{0}+\lvert\lambda_{1}\rvert\phi_{1}}{\lambda_{0}+\lvert\lambda_{1}\rvert}, \nonumber \\
\phi_{B}&=&\frac{\lambda_{2}\phi_{2}+\lvert\lambda_{1}\rvert\phi_{1}}{\lambda_{2}+\lvert\lambda_{1}\rvert}.
\label{eq:3}
\end{eqnarray}

In Eq.~(\ref{eq:1}) $\omega_{0}$ is the barrier height which separates the vapor at $\phi_{0}$ from the liquid at $\phi_{2}$, and is given by
\begin{eqnarray}
\omega_{0} &=& \frac{p}{2}\left(\phi_{1}-\phi_{0}\right)^{2}+\Delta\mu\frac{\phi_{1}-\phi_{0}}{\phi_{2}-\phi_{0}}, \nonumber \\
&=& \frac{q}{2}\left(\phi_{1}-\phi_{2}\right)^{2}+\Delta\mu\frac{\phi_{1}-\phi_{0}}{\phi_{2}-\phi_{0}}-\Delta\mu,
\label{eq:4}
\end{eqnarray}
where
\begin{eqnarray}
p &=& \frac{\lambda_{0}|\lambda_{1}|}{\lambda_{0}+|\lambda_{1}|}, \nonumber \\ 
q &=& \frac{\lambda_{2}|\lambda_{1}|}{\lambda_{2}+|\lambda_{1}|},
\label{eq:5}
\end{eqnarray} 
are constants determined from the compressibility $\lambda_{i}$.
By equating two formulae for $\omega_{0}$ in Eq.~(\ref{eq:4}), we obtain the location $\phi_{1}$ of the free energy barrier
\begin{equation}
\phi_{1}=\frac{(p\phi_{0}-q\phi_{2})+\sqrt{pq\left(\phi_{0}-\phi_{2}\right)^{2}-2\Delta\mu(p-q)}}{p-q},
\label{eq:6}
\end{equation}
which depends on the over-saturation $\Delta\mu>0$ for the oversaturated vapor and the undersaturation $\Delta\mu<0$ for the stretched liquid.  The vapor spinodal is defined when the metastable vapor phase at $\phi_{0}$ becomes unstable.  This is realized when $\phi_{1}=\phi_{0}$, which leads to the over-saturation for the gas spinodal
\begin{equation}
\Delta\mu_{gas}=\frac{q}{2}\left(\phi_{2}-\phi_{0}\right)^{2}
\label{eq:7}
\end{equation}
Similarly the liquid spinodal occurs when the undersaturation is given by
\begin{equation}
\Delta\mu_{\rm spin}=-\frac{p}{2}\left(\phi_{2}-\phi_{0}\right)^{2}
\label{eq:8}
\end{equation}
for the stretched liquid. Since we are interested in the critical bubble formation in the stretched fluid, we will consider the region from coexistence $\Delta\mu=0$ to the liquid spinodal $\Delta\mu_{\rm spin}<0$. In contrast to the previous models~\cite{Shen,Unger,Binder,Wilemski} where the compressibility diverges continuously as the spinodal is approached,  the compressibility remains finite until the spinodal point is reached in our triple-parabolic model as the curvature $\lambda_{0}$ and $\lambda_{2}$ is fixed.

\begin{figure}[htbp]
\begin{center}
\includegraphics[width=0.90\linewidth]{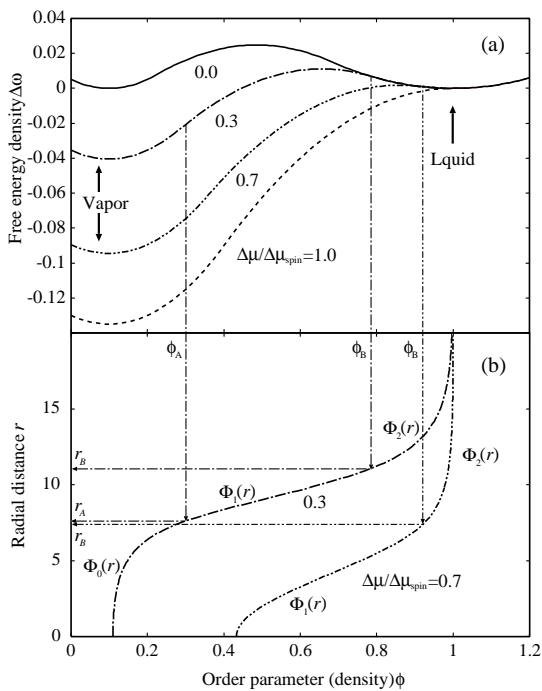}
\end{center}
\caption{
(a) The triple-parabolic free energy from CNT regime near the coexistence to the spinodal regime near the liquid spinodal for the case-I of section III (Table~\ref{tab:1}). (b) The corresponding critical bubble at the CNT regime ($\Delta\mu/\Delta\mu_{\rm spin}=0.3$) and at the spinodal regime ($0.7$).
}
\label{fig:1}
\end{figure}

In Fig.~\ref{fig:1}(a) we show typical shapes of the triple-parabolic free energy $\Delta \omega$ together with the corresponding critical bubble in Fig.~\ref{fig:1}(b).  The radii $r_{A}$ and $r_{B}$ are the matching radius that satisfies $\phi\left(r_{A}\right)=\phi_{A}$ and $\phi\left(r_{B}\right)=\phi_{B}$. Since the free energy consists of three parabolas corresponding to the vapor, spinodal and liquid parts, the density profile of the critical bubble consists of three parts that correspond to the three parts of free energy when $\Delta\mu/\Delta\mu_{\rm spin}=0.3$ near the coexistence.  However, as the undersaturation increases ($\lvert\Delta\mu\rvert$ becomes large) and it approaches the liquid spinodal $\Delta\mu_{\rm spin}$, the density profile consists of only the two parts correspond to the spinodal and the liquid parts.  We use the terminology "CNT regime" for the former regime where the classical nucleation theory (CNT) is expected to be qualitatively correct, and "spinodal regime" for the latter where the spinodal nucleation~\cite{Debenedetti,Unger,Binder,Wilemski} is expected to occur.


\subsection{Density Profile of Critical Bubble}

\subsubsection{CNT regime}
Density profile of spherically symmetric critical bubble can be obtained from the Euler-Lagrange equation $\delta W/\delta \phi=0$, which leads to the differential equation 
\begin{equation}
\frac{d^{2}\Phi_{i}}{dr^{2}}+\frac{2}{r}\Phi_{i}\pm \Gamma_{i}^{2}\Phi_{i}=0,\;\;\;i=0,1,2
\label{eq:9}
\end{equation}
for the three parabolas in Eq.~(\ref{eq:1}), where $\Gamma_{i}=\sqrt{\lvert\lambda_{i}\rvert/2c}$ and $\Phi_{i}(r)=\phi(r)-\phi_{i}$, and $+$ sign is used for $i=1$ and $-$ is used for $i=0, 2$ for $\pm$. These differential equations should be solved with boundary conditions:
\begin{eqnarray}
\Phi_{0}\left(r=r_{A}\right)&=&\Phi_{0A}=\phi_{A}-\phi_{0} \nonumber \\
\Phi_{1}\left(r=r_{A}\right)&=&\Phi_{1A}=\phi_{A}-\phi_{1} \nonumber \\
\Phi_{1}\left(r=r_{B}\right)&=&\Phi_{1B}=\phi_{B}-\phi_{1} \label{eq:10} \\
\Phi_{2}\left(r=r_{B}\right)&=&\Phi_{2B}=\phi_{B}-\phi_{2} \nonumber \\
\Phi_{2}\left(r\rightarrow \infty \right) &=& 0 \nonumber
\end{eqnarray}

together with the condition for $\Phi_{0}$ at the origin given by
\begin{equation}
\left.\frac{d\Phi_{0}}{dr}\right|_{r\rightarrow 0}=0
\label{eq:11}
\end{equation}
for the critical bubble.  In this CNT regime near the two-phase coexistence, both the matching radius $r_{A}$ and $r_{B}$ exist. When the under saturation $\Delta \mu (<0)$ is further increased, the matching radius $r_{A}$ approaches zero and disappears.  Then only the matching radius $r_{B}$ can exist.  We have named this regime near the spinodal point the spinodal regime.

The solution of this Euler-Lagrange equation in Eq.~(\ref{eq:9}) for the critical bubble are given by
\begin{eqnarray}
\Phi_{0}(r) &=& \Phi_{0A}r_{A} {\rm csch}\left(\Gamma_{0}r_{A}\right)\sinh\left(\Gamma_{0}r\right)/r
\nonumber \\
\Phi_{1}(r) &=& 
\csc\left(\Gamma_{1}\left(r_{A}-r_{B}\right)\right) \nonumber \\
&& \times (
-\Phi_{1B}r_{B}\sin\left(\Gamma_{1}\left(r-r_{A}\right)\right) \label{eq:12} \\
&&
+\Phi_{1A}r_{A}\sin\left(\Gamma_{1}\left(r-r_{B}\right)\right)
)/r \nonumber \\
\Phi_{2}(r) &=& \Phi_{2B}r_{B}\exp\left(-\Gamma_{2}r+\Gamma_{2}r_{B}\right)/r \nonumber
\end{eqnarray}
For the three parts $i=0, 1, 2$ respectively.  

Finally, the matching radii $r_{A}$ and $r_{B}$ are determined from the simultaneous equation
\begin{eqnarray}
\left.\frac{d\Phi_{2}}{dr}\right|_{r=r_{B}} &=& \left.\frac{d\Phi_{1}}{dr}\right|_{r=r_{B}}, \nonumber \\
\left.\frac{d\Phi_{1}}{dr}\right|_{r=r_{A}} &=& \left.\frac{d\Phi_{0}}{dr}\right|_{r=r_{A}},
\label{eq:13} 
\end{eqnarray}
where only $\Phi_{1}$ is the functions of both $r_{A}$ and $r_{B}$.  This simultaneous equation can be solved numerically using standard algorithms such as the Newton-Raphson method.

\subsubsection{Spinodal regime}

In this case $r_{A}$ becomes zero.   Therefore the $i=0$ part of the free energy density in Eq.~(\ref{eq:1}) and its solution $\Phi_{0}$ disappears.  Then the boundary condition Eqs.~(\ref{eq:10}) and (\ref{eq:11}) for $\Phi_{0}$ should be replaced by the same condition for $\Phi_{1}$.  The solution for the Euler-Lagrange equation for $\Phi_{2}$ is the same as Eq.~(\ref{eq:12}), but the one for $\Phi_{1}$ now reads
\begin{equation}
\Phi_{1}(r)=\Phi_{1B}r_{B}{\rm csc}\left(\Gamma_{1}r_{B}\right)\sin\left(\Gamma_{1}r\right)/r,
\label{eq:14}
\end{equation}
for the critical bubble in the spinodal regime.

In this case, the matching radius $r_{B}$ is simply determined from the equation
\begin{equation}
\left.\frac{d\Phi_{2}}{dr}\right|_{r=r_{B}} = \left.\frac{d\Phi_{1}}{dr}\right|_{r=r_{B}} 
\label{eq:15}
\end{equation}
which is explicitly written as
\begin{equation}
r_{B}\left(\lambda_{2}\Gamma_{1}\cot\left(\Gamma_{1}r_{B}\right)-\lvert\lambda_{1}\rvert\Gamma_{2}\right)=\lambda_{2}+\lvert\lambda_{1}\rvert,
\label{eq:16}
\end{equation} 
that does not depend on the understaturation $\Delta\mu$.  Therefore, the matching radius $r_{B}$ is constant in the spinodal regime.  Again this equation should be solved numerically.

In Fig.~\ref{fig:1}(b), we showed the typical density profiles in the CNT regime and in the spinodal regime.  The critical bubble is larger in the CNT regime than in the spinodal regime.  However, the density difference between the inside and the outside of the bubble becomes smaller in the spinodal regime than in the CNT regime.  Correspondingly, the interfacial thickness looks diffuse~\cite{Unger} as the spinodal is approached.

\subsection{Work of Formation of Critical Bubble}
\subsubsection{CNT regime}

Once we know the density profile of the bubble and cavity, it is straightforward to calculate the work of formation $W^{*}$ of the critical bubble.  To this end, we can use the formula
\begin{equation}
W^{*} =\int_{0}^{\infty}4\pi r^{2}\left(\Delta\omega-\frac{1}{2}\phi\frac{\partial \Delta\omega}{\partial \phi}\right)dr
\label{eq:17}
\end{equation}
derived by Cahn and Hilliard~\cite{Cahn2}.

Inserting the density profile Eq.~(\ref{eq:12}) in the CNT regime, we can calculate the integral in Eq.~(\ref{eq:17}) analytically, and we obtain
\begin{equation}
W=W_{0}+W_{1}+W_{2}
\label{eq:18}
\end{equation}
where
\begin{eqnarray}
W_{0}&=&\frac{4\pi}{3}r_{A}^{3}\Delta\mu + 2\pi\lambda_{0}\phi_{0}\frac{\left(1-\Gamma_{0}r_{A}\coth\left(\Gamma_{0}r_{A}\right)\right)\Phi_{0A}r_{A}}{\Gamma_{0}^{2}} \nonumber \\
W_{1}&=& \frac{4\pi}{3}\left(r_{B}^{3}-r_{A}^{3}\right)\left(-\Delta\mu\frac{\phi_{1}-\phi_{0}}{\phi_{2}-\phi_{0}}+\Delta\mu+\omega_{0}\right)  \nonumber \\
&&+\frac{2\pi|\lambda_{1}|\phi_{1}}{\Gamma_{1}^{2}}\left(-\Phi_{1A}r_{A}+\Phi_{1B}r_{B}\right. \label{eq:19} \\
&&+\left.\Gamma_{1}\left(\Phi_{1A}r_{A}^{2}+\Phi_{1B}r_{B}^{2}\right)\cot\left(\Gamma_{1}\left(r_{A}-r_{B}\right)\right)\right. \nonumber \\
&&-\left.\Gamma_{1}\left(\Phi_{1A}+\Phi_{1B}\right)r_{A}r_{B}{\rm csc}\left(\Gamma_{1}\left(r_{A}-r_{B}\right)\right)\right) \nonumber \\
W_{2} &=& -\frac{2\pi\lambda_{2}\phi_{2}\Phi_{2B}r_{B}}{\Gamma_{2}^{2}}\left(1+r_{B}\Gamma_{2}\right) \nonumber 
\end{eqnarray}
are the contributions of the three parts (vapor, spinodal, liquid) of the free energy density in Eq.~(\ref{eq:1}).  Note that $\Delta\mu<0$ for the critical bubble in the undersaturated liquid.


\subsubsection{Spinodal regime}

Since, we only have the solution $\Phi_{2}$ in Eq.~(\ref{eq:12}) and $\Phi_{1}$ given by Eq.~(\ref{eq:14}) for the critical bubble in the spinodal regime, we have 
\begin{equation}
W=W_{1}+W_{2}
\label{eq:20}
\end{equation}
where $W_{2}$ is given by Eq.~(\ref{eq:19}) but $W_{1}$ is given by
\begin{eqnarray}
W_{1} &=& \frac{4\pi}{3}r_{B}^{3}\left(-\Delta\mu\frac{\phi_{1}-\phi_{0}}{\phi_{2}-\phi_{0}}+\Delta\mu+\omega_{0}\right) \nonumber \\
&&+\frac{2\pi|\lambda_{1}|\phi_{1}\Phi_{1B}r_{B}}{\Gamma_{1}^{2}}\left(1-\Gamma_{1}r_{B}\cot\left(\Gamma_{1}r_{B}\right)\right)
\label{eq:21}
\end{eqnarray}
which can also be derived by setting $r_{A}=0$ in Eq.~(\ref{eq:19}).

\subsubsection{work of formation of classical critical bubble}

Here, we will summarize the standard formula for the work of formation of critical bubble from the classical nucleation theory (CNT)~\cite{Kelton,Oxtoby,Debenedetti}. Using the capillary approximation, the work of formation of critical bubble is given by the sum of volume term and the surface term:
\begin{equation}
W=-\frac{4\pi}{3}r^3\lvert\Delta\mu\rvert + 4\pi r^{2}\gamma,
\label{eq:22}
\end{equation}
where the critical bubble is assumed to be a sphere with the classical radius $r$ with the sharp interface of the macroscopic surface tension $\gamma$.  By maximizing this work of formation with respect to the bubble radius $r$, we obtain the critical radius $r_{\rm cl}$ of the critical bubble and its work of formation $W^{*}$ as
\begin{equation}
r_{\rm cl}=\frac{2\gamma}{\lvert\Delta\mu\rvert},\;\;\;\;\;W_{*}=\frac{16\pi\gamma^{3}}{3\Delta\mu^{2}}
\label{eq:23}
\end{equation}
where $\gamma$ should be calculated at the liquid-vapor coexistence when $\Delta\mu=0$:
\begin{eqnarray}
\gamma &=& 2\sqrt{c}\int_{\phi_{0}}^{\phi_{2}}\sqrt{\Delta \omega} d\phi \nonumber \\
       &=& \gamma_{0}+\gamma_{1}+\gamma_{2}
\label{eq:24}
\end{eqnarray}
which consists of three contributions from the three parts of the free energy density.  Due to the simple parabolic free energy for $\Delta \omega$, Eq.~(\ref{eq:24}) can be calculated analytically, and the results are~\cite{Granasy1}
\begin{eqnarray}
\gamma_{0}&=& \frac{1}{2}\sqrt{2\lambda_{0}c}\left(\phi_{A}-\phi_{0}\right)^{2}
\nonumber \\
\gamma_{1} &=& 2\sqrt{\lvert\lambda_{1}\rvert c}\left[\frac{x}{2}\sqrt{\xi^{2}-x^{2}}-\frac{\xi}{2}\arcsin\left(-\frac{x}{\sqrt{\xi}}\right)\right]_{x=\phi_{A}-\phi_{1}}^{x=\phi_{B}-\phi_{1}} 
\label{eq:24x} \\
\gamma_{2} &=& \frac{1}{2}\sqrt{2\lambda_{2}c}\left(\phi_{2}-\phi_{B}\right)^{2}
\nonumber
\end{eqnarray}
Equation (\ref{eq:23}) reduces to the standard expression of the CNT~\cite{Debenedetti,Oxtoby} if we recover the original definition $\Delta\mu$ in Eq.~(\ref{eq:2}) and replace $\Delta\mu$ by $\rho_{\rm l}\Delta\mu$  or by the pressure difference $\Delta P$.

\section{Numerical Results and Discussions}

\begin{figure}[htbp]
\begin{center}
\includegraphics[width=0.85\linewidth]{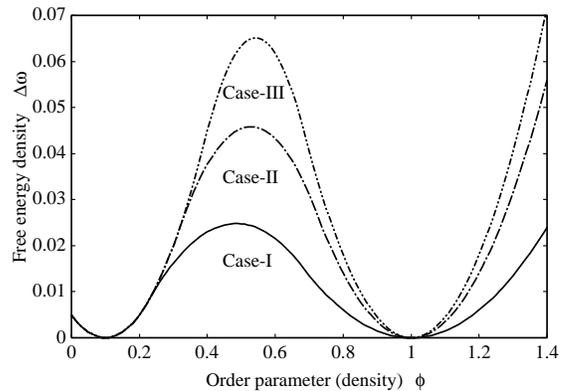}
\end{center}
\caption{
Three shapes of triple-parabolic free energy corresponding to the three sets of free energy parameter listed in Table ~\ref{tab:1} studied in this article at the two-phase coexistence.
}
\label{fig:2}
\end{figure}

In order to check the universality of the scaling relations for the various properties of critical bubble found by Shen and Debenedetti~\cite{Shen}, we use this square-gradient density-functional model with triple-parabolic free energy and study the various properties of the critical bubble.  We use several typical free energy parameters to check the universality.  Fig.~\ref{fig:2} shows the three shapes of triple-parabolic free energy used in our work.  Since the vapor is more compressible than the liquid in general, we choose $\lambda_{0}=1$ and $\lambda_{2}<\lambda_{0}$.  The three sets of the free energy parameters used are summarized in Table~\ref{tab:1}. Since our model is specified only by the compressibilities $\lambda_{i}$ and the undersaturation $\Delta\mu$, we will not consider the temperature effect~\cite{Shen,Kusaka} explicitly for it will come in through the temperature dependence and the interrelations between $\lambda_{i}$.

\begin{table}[htbp]
\caption{
Three sets of free energy parameters used in this work to check the universality of the scaling properties of the critical bubble.}
\label{tab:1}
\begin{center}
\begin{tabular}{c|cccccc}
\hline
model & $c$ & $\phi_{0}$ & $\phi_{2}$ &  $\lambda_{0}$ & $\lambda_{1}$ & $\lambda_{2}$ \\
\hline
case-I & 1.0 & 0.1 & 1.0 & 1.0 & -0.5 & 0.3 \\
case-II & 1.0 & 0.1 & 1.0 & 1.0 & -1.0 & 0.7 \\
case-III & 1.0 & 0.1 & 1.0 & 1.0 & -2.0 & 0.9 \\
\hline
\end{tabular}
\end{center}
\end{table}

\begin{figure}[htbp]
\begin{center}
\includegraphics[width=0.80\linewidth]{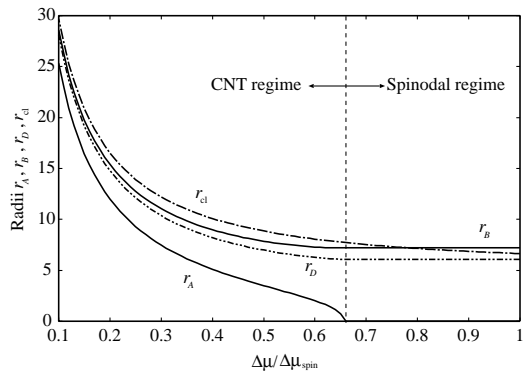}
\end{center}
\caption{
The matching radii $r_{A}$ and $r_{B}$ as well as the classical radius $r_{\rm cl}$ from the CNT for the case-I.  Also shown is the equimolar dividing radius $r_{D}$.  In the spinodal regime $r_{A}$ disappears, radii $r_{B}$ and $r_{D}$ become constant. 
}
\label{fig:3}
\end{figure}

Figure \ref{fig:3} shows the matching radius $r_{A}$ and $r_{B}$ as the functions of the scaled undersaturation $\Delta \mu/\Delta \mu_{\rm spin}$ for case-I. The matching radius $r_{A}$ becomes zero as the undersaturation enters the spinodal regime, while the matching radius $r_{B}$ survives and it becomes constant and independent of the undersaturation in the spinodal regime as predicted from Eq.~(\ref{eq:16}).  We also show the classical radius $r_{\rm cl}$ of the classical critical bubble predicted from Eq.~(\ref{eq:23}) of the CNT.

In Fig.~\ref{fig:3} we also show the equimolar dividing radius $r_{D}$ determined from
\begin{equation}
\int_{0}^{r_{D}}\left(\phi(r)-\phi_{\rm org}\right)4\pi r^2 dr
=\int_{r_{D}}^{\infty}\left(\phi_{2}-\phi(r)\right)4\pi r^2 dr
\label{eq:25}
\end{equation}
where
\begin{equation}
\phi_{\rm org}=\phi(r\rightarrow 0)
\label{eq:26}
\end{equation}
is the density at the origin of the bubble.  We use this definition Eq.~(\ref{eq:25}) of the dividing surface used by Shen and Debenedetti~\cite{Shen} though it is slightly different from the usual definition of the dividing surface where $\phi_{\rm org}$ should be replaced by $\phi_{0}$.  Since the density profile $\phi(r)$ is given by the analytical formulae in Eqs.~(\ref{eq:12}) and (\ref{eq:14}), not only the density at the origin $\phi_{\rm org}$ in Eq.~(\ref{eq:26}) but the integrals on the both side of Eq.~(\ref{eq:25}) are given by the analytic formula.  Then, the non-linear equation Eq.~(\ref{eq:25}) for the equimolar dividing radius $r_{D}$ can be solved numerically.  By a careful inspection of Eq.(\ref{eq:25}) in the spinodal regime, we notice that the dividing radius $r_{D}$ is constant as the matching radius $r_{B}$ in the spinodal regime.

From Fig.~\ref{fig:3} we notice that even though the classical radius $r_{\rm cl}$ is continuously decreasing as the undersaturation $\lvert\Delta\mu\rvert$ increases, the matching radius $r_{A}$ disappears while the matching radius $r_{B}$ and the dividing radius $r_{D}$ becomes constant in the spinodal regime up to the liquid spinodal $\Delta\mu_{\rm spin}$.  It is commonly believed that the radius of the critical nucleus diverges as the spinodal is approached~\cite{Cahn2,Unger}.  However, this conclusion is derived from some mathematical models and could be model-dependent.  In fact, this divergence is related to the third derivative of the free energy in the classical work of Cahn-Hilliard~\cite{Cahn2}.  Also, the conclusion of diverging critical nucleus drawn by Unger-Klein~\cite{Unger} is based on the third expansion of the free energy that contains up to $\phi^{4}$.  This third derivative does not exist in our triple-parabolic model.  Therefore, the extent of these anomalous divergences of the radius and the interfacial width of the critical nucleus depend strongly on the model used.  Of course, such an anomaly will not be observed in a real experiment as the meaning of nucleation becomes already obscure near the spinodal as the work of formation is of the order of the thermal energy $W_{*}\simeq k_{B}T$ near the spinodal~\cite{Binder}.

\begin{figure}[htbp]
\begin{center}
\includegraphics[width=1.00\linewidth]{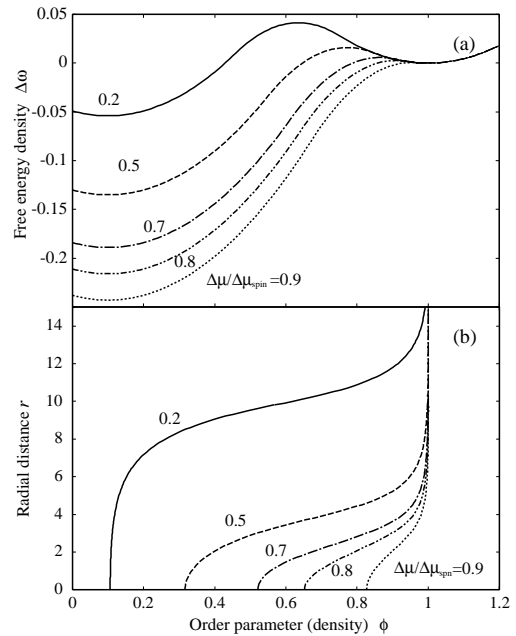}
\end{center}
\caption{
(a) The evolution of the triple-parabolic free energy from CNT regime to the spinodal regime for the case-III. (b) The corresponding evolution of the density profile of the critical bubble.  It seems that the interface of the bubble changes from sharp interface to diffuse one as the spinodal is approached. 
}
\label{fig:4}
\end{figure}

Figure \ref{fig:4}(a) shows the evolution of the free energy from the CNT regime near the two phase coexistence to the spinodal regime for the case-III.  The corresponding evolution of the density profile is shown in Fig.~\ref{fig:4}(b). Although the density profile seems to become flat and diffuse as the spinodal is approached, in fact the shape does not change even though the density difference between the inside and the outside of the bubble decreases as the positions of the dividing radius $r_{D}$ as well as the matching radius $r_{B}$ do not change in this spinodal regime.

\begin{figure}[htbp]
\begin{center}
\includegraphics[width=0.75\linewidth]{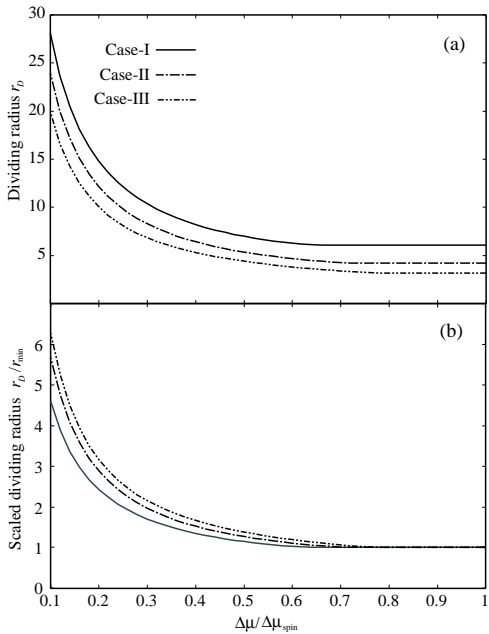}
\end{center}
\caption{
(a) The equimolar dividing radius $r_{D}$ as the functions of the scaled undersaturation $\Delta \mu/\Delta \mu_{\rm spin}$ for the three cases.  (b) The test of the scaling relation for the $r_{D}/r_{\rm min}$.  The three curves do not overlap in the CNT regime.  Scaling is recovered only in the spinodal regime.
}
\label{fig:5}
\end{figure}

The equimolar dividing radius $r_{D}$ as the function of the scaled undersaturation $\Delta \mu/\Delta \mu_{\rm spin}$ is shown in Fig.~\ref{fig:5}(a) and the scaled radius $r_{D}/r_{\rm min}$ are shown in Fig.~\ref{fig:5}(b), where $r_{\rm min}$ is the minimum of $r_{D}$ which is the values at the spinodal regime which is in fact the constant.  In contrast to the numerical results of Shen and Debenedetti\cite{Shen} for the Lennard-Jones fluid, the scaling relation for the dividing radius $r_{D}$ is barely satisfied in the CNT regime for the present triple-parabolic free energy model as three curves do not fall into a single curve as shown in Fig.~\ref{fig:5}(b). The scaling, however, recovered in the spinodal regime as $r_{D}$ is constant.  

Shen and Debenedetti~\cite{Shen} also noticed that the density at the origin $\phi_{\rm org}$
\begin{equation}
\phi_{\rm org}=\phi(r\rightarrow 0)
\label{eq:27}
\end{equation} 
as the function of scaled undersaturation $\Delta \mu/\Delta \mu_{\rm spin}$ is represented by a single curve, and satisfies the scaling. Furthermore, the curve shows shallow minimum at the CNT-spinodal boundary. They~\cite{Shen} argued that this anomaly is due to the change of the nucleation behavior from CNT-nucleation to the spinodal nucleation.  Similar anomalies are also found for various radii and material excess of the critical nucleus in the self-consistent-field (SCF) model of binary polymer blend.~\cite{Wood}

\begin{figure}[htbp]
\begin{center}
\includegraphics[width=0.80\linewidth]{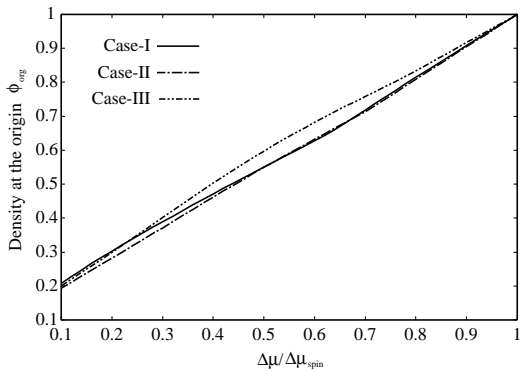}
\caption{
The density at the origin $\phi_{\rm org}$ as the functions of the scaled undersaturation $\Delta \mu/\Delta \mu_{\rm spin}$ for case-I to case-III.  Note that the bulk liquid density is $\phi_{2}=1$ and the bulk vapor density is $\phi_{0}=0.1$ (Table~\ref{tab:1}).
}
\end{center}
\label{fig:6}
\end{figure}

They~\cite{Shen} also found that the mean density $\phi_{\rm mean}$ which is defined as the average density calculated by integrating the bubble density $\phi(r)$ from the origin to the equimolar dividing radius $r_{D}$
\begin{equation}
\phi_{\rm mean} = \int_{0}^{r_{D}}\phi(r)4\pi r^{2}dr/\frac{4\pi r_{D}^{3}}{3}
\label{eq:28}
\end{equation}
also shows the scaling rule, and is expressed by a single curve as the function of the scaled undersaturation $\Delta \mu/\Delta \mu_{\rm spin}$.

Figure 6 shows the density at the origion $\phi_{\rm org}$ of our square-gradient model for the three cases.  In contrast to the results of Shen and Debenedetti~\cite{Shen} the scaling relation is marginally satisfied.  In particular, we cannot find out the anomalous density minimum found previously~\cite{Shen}.  In fact, since both calculations of ours and of Shen and Debenedetti is based on the density functional theory and is limited within the mean-field approximation, the anomalous spinodal nucleation is beyond the scope of our study as well as of Shen-Debenedetti's.  Thus the anomalous minimum found by them~\cite{Shen} is not due to the change of the style of nucleation to anomalous spinodal nucleation but merely due to the characteristics of the Lennard-Jones fluid.

\begin{figure}[htbp]
\begin{center}
\includegraphics[width=0.80\linewidth]{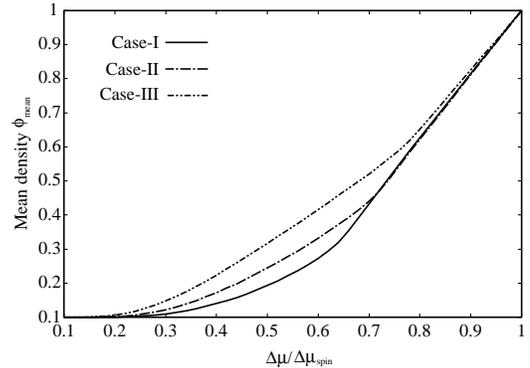}
\end{center}
\caption{
The mean density $\phi_{\rm mean}$ as the function of the scaled undersaturation $\Delta \mu/\Delta \mu_{\rm spin}$.
}
\label{fig:7}
\end{figure}

Figure \ref{fig:7} shows the mean density $\phi_{\rm mean}$ of the bubble as the function of the scaled undersaturation $\Delta \mu/\Delta \mu_{\rm spin}$.  Again, the scaling relation is marginally satisfied in the CNT regime in contrast to the numerical results by Shen and Debenedetti~\cite{Shen}.  However, the scaling is recovered in the spinodal regime, which is due to the facts that the equimolar dividing radius $r_{D}$ becomes constant and the density profile becomes universal curve in the spinodal regime which will be discussed later in this section.  

\begin{figure}[htbp]
\begin{center}
\includegraphics[width=0.80\linewidth]{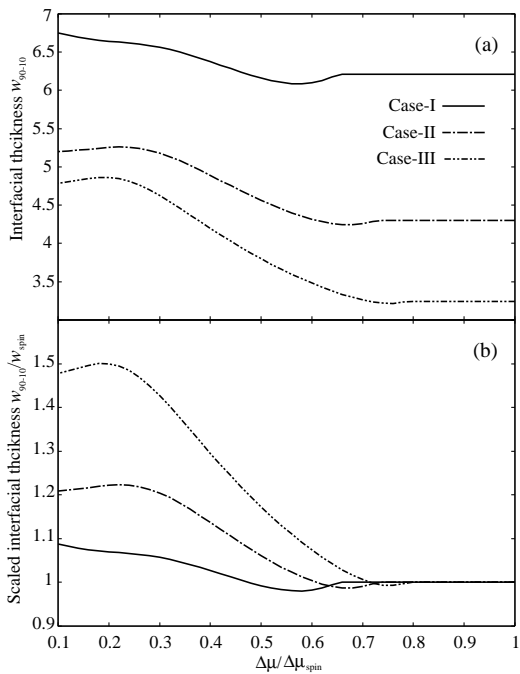}
\end{center}
\caption{
(a) The interfacial thickness calculated from $w_{90-10}=r_{90}-r_{10}$ as the functions of the scaled undersaturation $\Delta \mu/\Delta \mu_{\rm spin}$. (b) The scaled thickness $w_{90-10}/w_{\rm spin}$ divided by the thickness in the
spinodal regime.
}
\label{fig:8}
\end{figure}

It has been argued for many years~\cite{Debenedetti,Binder,Cahn2,Unger} that the interfacial thickness of the critical nucleus diverges as the spinodal is approached.  Therefore, the nucleation becomes anomalous near the spinodal as the critical nucleus becomes ramified fractal object~\cite{Unger} rather than the compact spherical droplet.  Shen and Debenedetti~\cite{Shen} calculated the interfacial thickness of the critical bubble of the Lennard-Jones fluid defined by the 10-90 width $w_{90-10}=r_{90}-r_{10}$ through
\begin{equation}
\phi_{2}-\left(\phi_{2}-\phi_{\rm org}\right)\times 0.1 = \phi\left(r_{90}\right)
\label{eq:29}
\end{equation}
where $\phi_{\rm org}$ is defined in Eq.(\ref{eq:27}), and
\begin{equation}
\phi_{\rm org}+\left(\phi_{2}-\phi_{\rm org}\right)\times 0.1 =\phi\left(r_{10}\right). 
\label{eq:30}
\end{equation}
They found that the thickness $w_{90-10}$ of the Lennard-Joes fluid does diverge as the spinodal is approached.  Furthermore, the thickness of the Lennard-Jones fluid exhibits scaling.  By dividing minimum thickness, the interfacial thickness is represented by a single universal curve if it is plotted against the scaled undersaturation $\Delta \mu/\Delta \mu_{\rm spin}$.

We showed the interfacial thickness of our square-gradient density-functional model derived from $w_{90-10}=r_{90}-r_{10}$ in Fig.~\ref{fig:8}(a) as the function of the scaled undersaturation $\Delta \mu/\Delta \mu_{\rm spin}$.  By careful examination of Eqs.~(\ref{eq:29}) and (\ref{eq:30}) with Eqs.~(\ref{eq:12}) and (\ref{eq:14}), we can easily prove that both the radius $r_{90}$ and $r_{10}$ are constant in the spinodal regime.  Then, so is the interfacial thickness $w_{90-10}$. 

In fact, the thickness $w_{90-10}=r_{90}-r_{10}$ becomes constant in the spinodal regime but the magnitude differs from case-I to case-III in Fig.~\ref{fig:8}(a).  The scaling relation is also tested in Fig.~\ref{fig:8}(b) by dividing the interfacial thickness $w_{90-10}$ by the width $w_{\rm spin}$ in the spinodal regime.  The three distinct curves in Fig.~\ref{fig:8}(b) show that the scaling for the interfacial thickness is strongly violated in the CNT regime in contrast to the results of Shen and Debenedetti~\cite{Shen} for the Lennard-Jones fluid.  The scaling, however, is recovered in the spinodal regime as the thickness $w_{90-10}=w_{\rm spin}$ becomes constant.

\begin{figure}[htbp]
\begin{center}
\includegraphics[width=0.80\linewidth]{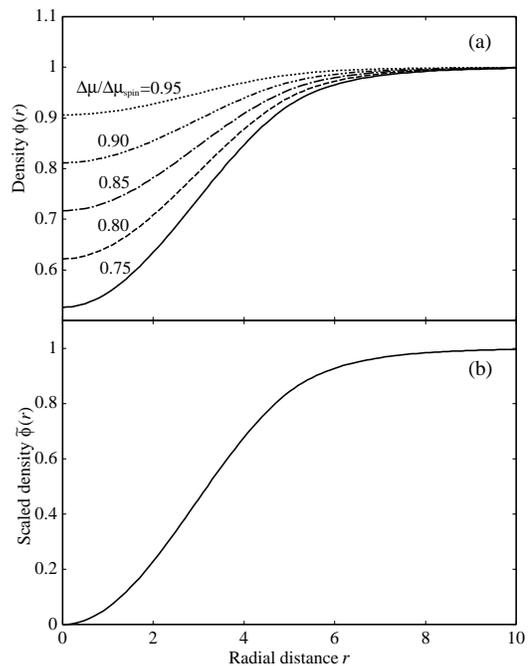}
\end{center}
\caption{
(a) The density profile $\phi(r)$ of the critical bubble for the several values of the undersaturation $\Delta\mu$ in the spinodal regime for Case-II.  (b) The corresponding scaled density profile $\tilde{\phi}(r)$ defined by Eq.~(\ref{eq:31}). All curves in (a) overlap into a single universal curve.
}
\label{fig:9}
\end{figure}

In Fig.~\ref{fig:9}(a) we show the density profiles of critical bubble for several values of undersaturation $\Delta\mu$ in the spinodal regime.  The density profile changes continuously and it becomes flat and diffuse as the spinodal is approached as the density difference between the inside and the outside of the bubble becomes small.  The fact is that the interfacial thickness is constant as the radius $r_{90}$ and $r_{10}$ are constant in this regime, which can be further confirmed by inspecting the scaled density profile $\tilde{\phi}(r)$ defined by
\begin{equation}
\tilde{\phi}(r)=\frac{\phi(r)-\phi_{\rm org}}{\phi_{2}-\phi_{\rm org}},
\label{eq:31}
\end{equation}
which can be shown to become a universal curve independent of undersaturation $\Delta\mu$ in the spinodal regime:
\begin{eqnarray}
\tilde{\phi}(r) &=& \tilde{\phi}_{0}\lambda_{2}r_{B}{\rm csc}\left(\Gamma_{1}r_{B}\right)\left(\frac{\sin\left(\Gamma_{1}r\right)}{r}-\Gamma_{1}\right) \nonumber \\
&=& 1-\tilde{\phi}_{0}\lvert\lambda_{1}\rvert r_{B}\frac{\exp\left(-\Gamma_{2}r+\Gamma_{2}r_{B}\right)}{r}
\label{eq:32}
\end{eqnarray}
where
\begin{equation}
\tilde{\phi}_{0}^{-1} = \lvert\lambda_{1}\rvert+\lambda_{2}-\lambda_{2}r_{B}{\rm csc}\left(\Gamma_{1}r_{B}\right)\Gamma_{1}
\label{eq:33}
\end{equation}
which does not change the shape as the matching radius $r_{B}$ does not depend on the undersaturation $\Delta\mu$ in the spinodal regime.

We show in Fig.~\ref{fig:9}(b) the universal scaled density profile in the spinodal regime.  It is clear from this figure that the interfacial width is constant up to the spinodal in the spinodal regime.

Cahn and Hilliard~\cite{Cahn2} used the radius $r_{1/2}$ defined by
\begin{equation}
\left(\phi_{\rm org}+\phi_{2}\right)/2 =\phi\left(r_{1/2}\right)
\label{eq:34}
\end{equation}
as the typical size of the critical nucleus, and has found that $r_{1/2}$ diverges as the spinodal approached.  Similar divergence for $r_{1/2}$ was found for other models~\cite{Wilemski,Wood}.  However, it is clear from Fig.~\ref{fig:9}(b) that $r_{1/2}$ of our triple-parabolic feee energy model does not diverge but remains constant similar to $r_{B}$ and $r_{D}$ in the spinodal regime.

\begin{figure}[htbp]
\begin{center}
\includegraphics[width=0.77\linewidth]{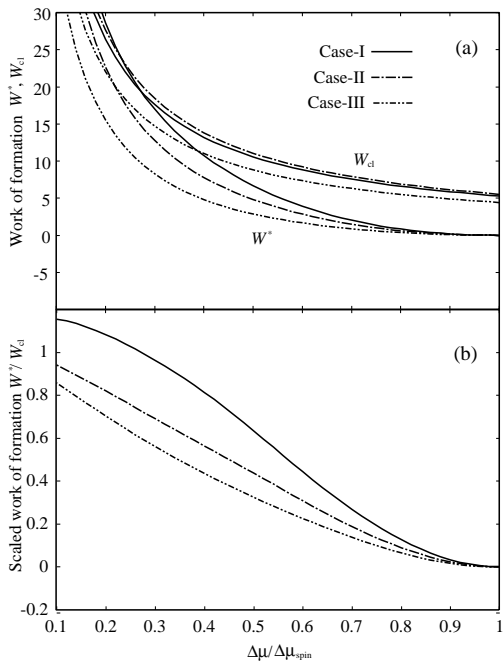}
\end{center}
\caption{
(a) The reversible work of formation $W^{*}$ of the critical bubble compared with the classical work of formation $W_{\rm cl}$ as the functions of the scaled
undersaturation $\Delta \mu/\Delta \mu_{\rm spin}$. (b) The scaled work of formation $W_{*}/W_{\rm cl}$ which does not exhibit the scaling property. 
}
\label{fig:10}
\end{figure}

Therefore, the diverging critical nucleus and the diverging interfacial thickness and corresponding fractal critical nucleus predicted originally by Cahn and Hillirad~\cite{Cahn2} and confirmed by others~\cite{Unger,Wood,Shen} cannot be applied to the critical bubble of our model.  Using the nucleation theorem~\cite{Viisanen,Oxtoby3}, Wilemski and Li~\cite{Wilemski} showed that the excess number of molecule
\begin{equation}
\Delta g = 4\pi \int_{0}^{\infty}\left[\phi_{2}-\phi(r)\right]r^{2}dr,
\label{eq:34a}
\end{equation}  
which is in fact the missing number of molecule in the bubble nucleation, is proportional to the isothermal compressibility $\kappa$ of the metastable liquid
\begin{equation}
\Delta g \propto \kappa.
\end{equation}
Since the isothermal compressibility $\kappa$ diverges as the spinodal is approached in the square-gradient density functional model with the polynomial free energy containing up to $\phi^{4}$ ($\phi^{4}$-field model)~\cite{Cahn2,Unger} or the Lennard-Jones liquid~\cite{Shen,Wilemski}, the corresponding $\Delta g$ diverges and so does the size of critical nucleus.  Now, in our triple-parabolic free energy model, the isothermal compressibility given by $\kappa^{-1}=\phi_{2}^{2}\lambda_{2}$ stays constant as the spinodal is approached, which is the reason why the size of bubble represented by the width $w_{90-10}$ of the interface does not diverge.  Therefore our triple-parabolic free-energy model that is similar but is slightly different from $\phi^{4}$-field model does not give the diverging interfacial thickness as the spinodal is approached.  Therefore, this type of parabolic model~\cite{Iwamatsu1,Bagdassarian,Barrett} may not be suitable to study the structure of nucleus near the spinodal.  This divergence at the spinodal is caused by the divergence of the isothermal compressibility of the metastable liquid as the spinodal is approached~\cite{Kalikmanov}.  However, the degree of this divergence should depend strongly on the mathematical structure of the model used, or the physical system experimentally studied.  In fact, in contrast to the prediction of Cahn and Hilliard~\cite{Cahn2}, the experimental results by Lefebvre et al.~\cite{Lefebvre} for polymer mixtures indicates that the size of critical nucleus monotonically decrease rather than diverges with increasing quench depth, and it remains finite at the spinodal.

Finally, Fig.~\ref{fig:10}(a) shows the reversible work of formation of the critical bubble calculated from Eqs.~(\ref{eq:19}) and (\ref{eq:21}) compared with the classical work of formation Eq.(\ref{eq:23}).  Our square-gradient density-functional model correctly predicts that the work of formation vanishes at the spinodal while the classical work remains finite at the spinodal.  The scaled work of formation $W^{*}/W_{\rm cl}$ barely shows scaling properties
\begin{equation}
\frac{W_{*}}{W_{\rm cl}}=1-\left(\frac{\Delta\mu}{\Delta\mu_{\rm spin}}\right)^{2}
\label{eq:35}
\end{equation}
proposed by Talanquer~\cite{Talanquer} and Shen and Debenedeti~\cite{Shen} in contrast to the results for the Lennard-Jones fluid~\cite{Shen}.  The shape changes from convex up for the Case-I, which is similar to previous work~\cite{Shen,Kusaka} for the Lennard-Jones and the square-well fluids, to the convex down for the Case-III, which is simlar to the results by M\"uller {\it et al.}~\cite{Muller} for the density fucntional culculation of bubble nucleation in the polymer mixture. Therefore, the various scaling rules~\cite{McGraw,Talanquer,Kashchiev,Kusaka,Kalikmanov} for the work of formation could depend strongly on the model the authors has chosen.  Then the scaling rule, such as Eq.~(\ref{eq:35}) may not always be valid for any model or any physical systems.  A similar modest scaling to ours was found by Kusaka~\cite{Kusaka}.

Although we do not obtain anomalous structural properties of the critical bubble near the spinodal, the work of formation of the critical bubble correctly vanishes at the spinodal.  Therefore anomalous spinodal nucleation~\cite{Debenedetti,Unger,Binder} should occur even within our model as soon as the work of formation is of the order of the thermal energy $W^{*}\simeq k_{B}T$. 

\section{Conclusion}

In this study, we use a square-gradient density-functional model with triple-parabolic free energy, to study the various properties of the critical bubble of homogeneous bubble nucleation.  In particular, we pay attention to the scaling properties of the reversible work of formation of critical bubble as well as other quantities such as the density at the center of critical bubble or the interfacial thickness. In contrast to the similar density-functional study of Shen and Debenedetti~\cite{Shen} who have found that not only the work of formation but other various quantities related to the critical bubble shows scaling rule, we found marginal scaling relation for our model.  Since our model is specified only by the compressibility and the undersaturation, our result seems to indicate that the liquid compressibility changes concomitantly with the vapor compressibility as the temperature is changed in the Lennard-Jones fluid, which would lead to the scaling rule found by Shen and Debenedetti~\cite{Shen} for different temperatures.  

Furthermore, we found that the structural anomalies of the critical bubble near the spinodal found in the $\phi^{4}$-field theory\cite{Unger,Binder} does not exist. Specifically we did not find anomalous minimum in the density at the center of critical bubble found by Shen and Debenedetti\cite{Shen} for the Lennard-Jones fluid when the undersaturation is increased toward the spinodal.  They~\cite{Shen} attributed this anomaly due to the spinodal nucleation.  We also found that the interfacial thickness of the critical bubble in the spinodal regime does not diverge near the spinodal as Unger and Klein~\cite{Unger} predicted. In contrast, the thickness remains constant up to the spinodal.  Furthermore, the interfacial profile remains the same in the spinodal regime as we scale the profile.  Therefore a part of the anomalous behavior of various quantities near the spinodal predicted by various authors using various different models could be largely model-dependent.  Such anomalies and their extent could partly be due to the mathematical structure of $\phi^{4}$-field or the Lennard-Jones fluid. Further theoretical study to test the scaling properties of the critical nucleus and bubble, in particular near the spinodal using other realistic models will be certainly necessary.  The critical cavity~\cite{Punnathanam2,Uline} in the stretched fluid can also be studied using this square-gradient density-functional model with triple-parabolic free energy.  Finally, the dynamics of the bubble formation will also be interesting, for that purpose numerically efficient cell dynamics system~\cite{Iwamatsu4} could be used.   

\begin{acknowledgments}
The author is grateful to Professor D. S. Corti (Purdue Univesity) for his useful comment on earlier version of this manuscript and valuable information about the theoretical works by Kusaka and the experimental results by Balsara's group in Berkeley.
\end{acknowledgments}

\end{document}